\documentclass[12pt,psfig]{article} \textheight=23 true cm \textwidth=16 true cm
\usepackage{psfig}
\usepackage{graphicx}

\begin{document}

\begin{center}
{\Large\bf Accelerating Universe as Window for Extra Dimensions
}\\[15 mm]
D. Panigrahi\footnote{Relativity and Cosmology Research Centre,
Jadavpur University, Kolkata - 700032, India , e-mail:
dibyendupanigrahi@yahoo.co.in , Permanent Address :
 Kandi Raj College, Kandi, Murshidabad 742137, India},
 Y. Z. Zhang\footnote{Permanent Address :
 Institute of theoretical Physics, Chinese Academy of Sciences,
 P.O.B. 2735, Beijing, China, e-mail: yzhang@itp.ac.cn}
  and S. Chatterjee\footnote{Relativity and Cosmology Research Centre, Jadavpur University,
Kolkata - 700032, India, e-mail : chat\_ sujit1@yahoo.com\\Correspondence to : S. Chatterjee} \\[10mm]

\end{center}

\begin{abstract}
Homogeneous  cosmological solutions are obtained in five
dimensional space time assuming equations of state $ p = k\rho $
and $ p_{1}= \gamma\rho$ where p is the isotropic 3 - pressure and
$p_{1}$, that for the fifth dimension. Using different values for
the constants k and $\gamma$ many known solutions are
rediscovered. Further the current acceleration of the universe has
led us to investigate higher dimensional gravity theory, which is
able to explain acceleration from a theoretical view point without
the need of introducing  dark energy by hand. We also extend a
recent work of Mohammedi where using a special form of the extra
dimensional scale factor a new interpretation of the higher
dimensional equations of motion is given and the concept of an
\emph{effective} four dimensional pressure is introduced.
Interestingly the 5D matter field remains regular while the
\emph{effective} negative pressure is responsible for the
inflation. Relaxing the assumptions of two equations of state we
also present a class of solutions which provide early deceleration
followed by a late acceleration in a unified manner. Relevant to
point out that in this case our cosmology apparently mimics the
well known quintessence scenario fuelled by a generalised
Chaplygin-type of fluid where a smooth transition from a dust
dominated model to a de Sitter like one takes place. Depending on
the relative magnitude of the different constants appearing in our
solutions we show that some of the cases are amenable to the
desirable property of dimensional reduction.
\end{abstract}
\bigskip
\section*{1. Introduction}
 There are growing evidences today that the current expansion of
 the universe is accelerating. It follows directly from the
 findings of Ia Supernovae  and indirectly from CMBR
 fluctuations. The latter observation points to the fact
 that the average mass density of the universe is very close to
 critical density. But the large scale structure of our universe
 indicates that normal gravitating ( but invisible ) matter can
 account for only 30\% into the energy budget. One is naturally
 left with remaining 70\%  of the energy which is some mysterious
 agent responsible for the cosmic acceleration. If we put faith in
 FRW type of models  then General Relativity is unambiguous about
 the need for some sort of dark energy source to explain the
 acceleration, which should behave like a fluid with a large
 negative pressure in the form of a time dependent cosmological
 constant or an evolving scalar field called \emph{quintessence}. However
 none of the existing dark energy models is completely
 satisfactory. Moreover, it is very difficult to construct a
 theoretical basis for the origin of this exotic matter, which is
 seen precisely at the current epoch when one needs the source for
 cosmic acceleration ( coincidence problem ).\\
 So there has been a resurgence of interests among relativists,
   field theorists, astrophysicists and people doing astroparticle
   physics both at theoretical and experimental levels to address
   the problems coming out of the recent extra galactic observations
   (for a lucid and fairly exhaustive exposition of some of these
   ideas one is referred to \cite{pd} and references therein ) without involving a
   mysterious form of scalar field by hand but looking for
   alternative approaches \cite{car},\cite{kuan} based on sound physical
   principles.Viswakarma \cite{vis} in a series of papers argued
   that it is possible to explain the recent observational
   findings in the frame work of a decelerating model also.
   Another suggestion is that light emitted from a distant
   supernova encounters an obstacle enroute to us and gets
   partially absorbed  apparently dimming the supernovae
   \cite{kalop} due to flavour oscillations. It occurs when
    there are several degrees of freedom whose interaction eigenstates
    coincide with the propagation eigenstates. Such particles can turn into
    other particles and evade detection. Other alternatives include
     modification of the Einstein- Hilbert action through the introduction
      of additional curvature terms, $R^{m}+ R^{n}$ ( $ m>0, n<0 $ and not
   necessarily integer) in the Lagrangian \cite{nkd},\cite{ua}. The effective Friedmann
   equations  contain extra terms coming from higher curvatures which may
    be viewed as a fluid, responsible for the current acceleration. However the
    resulting field equations are extremely difficult to solve and
    moreover, the cosmology is mostly unstable against
    perturbations. Hence this curvature quintessence has also of late somewhat
     fallen from grace.\\
   On the other hand , serious attempts are recently being made \cite{sc},\cite{bp},\cite{li}
    to incorporate the phenomenon of accelerating universe within the framework of higher
   dimensional space time itself without involving any mysterious scalar field
   with large negative pressure by hand. The attempt to unify gravity with other
   forces in nature is an active field of research. Some earlier
   works \cite{wes},\cite{cbh} have been directed at studying theories in which the
   dimensions of space time is greater that the ( 3+1 ) of the
   world we observe. Moreover, the advance of super gravity in 11D
   and super string in 10D indicate that the multidimensional
   space is apparently a fairly adequate reflection of dynamics of
   interaction over distances $ r\ll 10^{-16}$ cm. where unification
   of all types of forces may occur. Recent spurt in activities also
   stems from its applications to brane cosmology. The
  realisation that the universe is currently undergoing an
   accelerated expansion phase and the quest for the nature of the
   quintessence field have renewed the interest in higher
   dimensional gravity and their relation to cosmology. This is
   due to the fact that the higher dimensional corrections to the
   Einstein's field equations can be viewed as an effective fluid
   and this fluid can emulate the action of the homogeneous part
   of the quintessence field. Hence, in this extra dimensional
   quintessence scenario, what we observe as a new
   component of cosmic energy density is an effect of higher
   dimensional corrections to the  Einstein-Hilbert action. This
   approach has definite advantage over the standard quintessence
   scenario because we do not need to search for the quintessence
   scalar field and pick them by hand. On the contrary the
   extra fluid responsible for the acceleration is geometrical in
   origin having strong physical foundation and also in line with the spirit
   of general relativity as proposed first by Einstein \cite{einstein} and later developed
   by Wesson \cite{wes}. In a recent communication \cite{milton} it is also argued
   that quantum fluctuations in 4D spacetime do not give rise to dark energy.
   Rather a possible source of dark energy is the fluctuations in quantum fields
   , including quantum gravity inhabiting extra compactified dimensions.\\
   Here we have taken a 5D homogeneous line element with a
   zero curvature  spherically symmetric 3D space. The motivation for the
   present work is primarily twofold. Assuming two equations of state as
    $ p=k\rho$ and $p_{1}=\gamma\rho$ we have solved the field equations
   and in the process  have recovered some of the important earlier works
   in this field as special cases. On the other hand to conform to
   the current accelerating phase of the universe we have searched
   for quintessential behaviour of our solutions, if any. We get
   the interesting result in sections 2 and 3 that for a realistic 5D matter field
   characterised by $0< k$, $\gamma <1$ it is impossible to get accelerating
  universe with a power law solution for the 3D scale factor. This is in line
   with the 4D cosmology also. We also calculate the  limiting values of
$k$ and $\gamma$ when the cosmology shows the desirable property
of dimensional reduction. Moreover, while working in higher
dimensional theories it is not enough to show  spontaneous
compactification occurs. The cosmological consequences of the
shrinking extra dimensions should also be taken into account. In
this context we have extended a recent work of Mohammedi\cite{mh}
who gave an alternative interpretation of additional higher
dimensional terms appearing in the field equations to argue that a
regular matter field in higher dimension can generate, in
principle at least, an effective pressure which may be negative to
trigger an inflation of the 4D spacetime. This is dealt with in
section 4 where we have put forward a specific solution to
illustrate our point. In section 5 we have assumed a specific form
of the deceleration parameter to find a solution of the scale
factor for shear free expansion. We get a class of solutions with
interesting physical properties. An additional free parameter
appearing in the expression of the scale factor characterises the
form of the matter field similar to the well known form of the
generalised Chaplygin gas for quintessential models. The resulting
energy momentum tensor behaves like a mixture of cosmological
constant and a perfect fluid obeying higher dimensional equation
of state. When the cosmological radius is small the matter field
in the form of dust( for example) predominates giving a
decelerating expansion till the cosmological term takes over
effecting a smooth transition to the current accelerating phase,
while in the intermediate stage our cosmology interpolates between
different phases of the universe. This phenomena has been
exhaustively discussed in the context of quintessence in 4D
spacetime. However we are not aware models of similar kind in
higher dimensional spacetime, that too without assuming by hand
any form of an extraneous scalar field with mysterious properties
The paper ends with a discussion in section 6.

\section*{2. The Field Equations and its integrals}
We begin with considering a (d+4)-dimensional line-element
\begin{eqnarray}
  ds^{2} &=&
  dt^{2}-R^{2}\left(\frac{dr^{2}}{1-Kr^{2}}+r^{2}d\theta^{2}+
  r^{2}sin^{2}\theta d\phi^{2}\right) - A^{2}\gamma_{ab}dy^{a}dy^{b}
\end{eqnarray}
where $y^{a}$( a,b = 4,....        ,3+d) are the extra dimensional
coordinates and the 3D and extra dimensional scale factors R and A
depend on time only  and K is the 3D curvature and the compact
manifold is described by the metric $\gamma_{ab}$. For our
manifold $M^{1}\times S^{3}\times S^{d}$ the symmetry group of the
spatial section is $O(4) \times O(d+1) $. The stress tensor whose
form will be dictated by Einstein's equations must have the same
invariance leading to the energy momentum tensor as \cite{rd}
\begin{equation}
T_{00}=\rho~,~~T_{ij}= -p(t)g_{ij}~,~~ T_{ab}=-p_{d}(t)g_{ab}
\end{equation}
where the rest of the components vanish. Here p is the isotropic
3- pressure and $p_{d}$, that in the extra dimensions. Assuming
two equations of
 state $p=k\rho$ and $p_{d}=\gamma\rho$, we get from the Bianchi
identity $T^{AB}_{;A}=0$

\begin{equation}
\dot{\rho}+\left[3\frac{\dot{R}}{R}
(1+k)+d\frac{\dot{A}}{A}(1+\gamma)\right ]\rho =0
\end{equation}

The last equation integrates to
\begin{equation}
\rho = R^{-3(1+k)}A^{-d(1+\gamma)}
\end{equation}
Using equation(4) the independent field equations  for our metric
(1) are

\begin{eqnarray}
\frac{\rho}{2\beta} &=& \left\{3\frac{\dot{R^{2}}+ K}{R^{2}}-
\lambda \right\} + \frac{1}{2}d(d-1)\frac{\dot{A^{2}}}{A^{2}}
+3d\frac{\dot{R}\dot{A}}{RA}+ \frac{d}{2}(d-1)\frac{\kappa}{A^{2}}
\nonumber\\\nonumber
 &=& R^{-3(1+k)}A^{-d(1+\gamma)}\\
  &\equiv&  \left\{3\frac{\dot{R^{2}}+ K}{R^{2}}- \lambda\right \}
 +\frac{\rho_{0}}{2\beta}\\
 \frac{p}{2\beta} &=&
 \left\{-2\frac{\ddot{R}}{R}-\frac{\dot{R^{2}}}{R^{2}}-
 \frac{K}{R^{2}}+\lambda \right\}- d \frac{\ddot{A}}{A} -
 \frac{1}{2}d(d-1)\frac{\dot{A^{2}}}{A^{2}}-2d\frac{\dot{R}}{R}\frac{\dot{A}}{A}
 - \frac{d}{2}(d-1)\frac{\kappa}{A^{2}} \nonumber\\&=& kR^{-3(1+k)}A^{- d
 (1+\gamma)}\nonumber\\
 &\equiv&
  \left\{-2\frac{\ddot{R}}{R}-\frac{\dot{R^{2}}}{R^{2}}-
 \frac{K}{R^{2}}+\lambda \right\} + \frac{p_{0}}{2\beta}\\
\frac{p_{d}}{2\beta}  &=& - 3 \frac{\ddot{R}}{R} -
3\frac{\dot{R^{2}}}{R^{2}} - 3\frac{K}{R^{2}} + \lambda -
(d-1)\frac{\ddot{A}}{A} -
\frac{1}{2}(d-1)(d-2)\frac{\dot{A^{2}}}{A^{2}} \nonumber\\ &-&
3(d-1)\frac{\dot{R}}{R}\frac{\dot{A}}{A} -
\frac{1}{2}(d-1)(d-2)\frac{\kappa}{A^{2}} \nonumber\\
&=& \gamma R^{-3(1+k)}A^{- d (1+\gamma)}
 \end{eqnarray}
where $\lambda$ is a higher dimensional cosmological constant,
$\kappa$ is the curvature of the extra space and 2$\beta$ is the
$(d+4)$-dim. gravitational coupling constant. In what follows we
assume for simplicity a 5D spacetime$(d=1)$ although we believe
many of our results can be extended even when $d > 1$. However, in
section 4 we will have occasion again to discuss the general line
element(1). The last three equations are not independent. We take
the following two combinations  as our key equations to be solved
\begin{equation}
\frac{\ddot{R}}{R}+ \frac{(1+3k)}{2}\frac{(\dot{R^{2}}+ K
)}{R^{2}}+
\frac{(2+3k)}{2}\frac{\dot{R}\dot{A}}{RA}+\frac{1}{2}\frac{\ddot{A}}{A}= 0\\
\end{equation}
and\\
\begin{equation}
 \frac{\ddot{R}}{R}+(1+\gamma)\frac{\dot{R^{2}}+ K}{R^{2}}+
\gamma\frac{\dot{R}\dot{A}}{RA} =0
\end{equation}
At this stage  we take for simplicity K = 0, and the equation (9)
then yields a first integral as
\begin{equation}
A =
\frac{\alpha}{\dot{R^{\frac{1}{\gamma}}}R^{\frac{(1+\gamma)}{\gamma}}}
\end{equation}
where $\alpha$ is a constant of integration. For simplicity we
take a power law solution for the scale factor
\begin{equation}
R = t^{m}
\end{equation}
which gives via equation (10)
\begin{equation}
A= t^{\frac{-(2m+m\gamma-1)}{\gamma}}\\
\end{equation}
The equation (8) further restricts the value of m as
  $m=\frac{1}{2}~, ~~\frac{1-\gamma}{\gamma^{2}-3k\gamma+2}$.
The former  value of m gives
    $R=t^{\frac{1}{2}}$ and $A=t^{-\frac{1}{2}}$ as also
    $p=p_{5}=\rho=0$. Incidentally this is the wellknown solution
    of Chodos and Detweiler \cite{ch}  for a matter free 5D model.
The second value of m gives the following    solutions as
    \begin{eqnarray}
      R &=& t^{\frac{1-\gamma}{\gamma^{2}-3k\gamma+2}} \\
      A &=& t^{\frac{2\gamma-3k+1}{\gamma^{2}-3k\gamma+2}} \\
      \rho &=& \frac{3(1-\gamma)(\gamma-3k+2)}{(\gamma^{2}-3k\gamma+2)^{2}t^{2}} \\
      p &=& k\rho \\
      p_{5} &=& \gamma \rho
    \end{eqnarray}
\section*{3. Dynamical Behaviour }
In what follows we shall see that physical considerations put some
restrictions on the values of k and $\gamma$. If we believe in an
expanding universe $\gamma=1$ is clearly ruled out from equation (13). Further  it is
evident that the 3- space expands for $0<\gamma<1$ and $0<k<1$.
Again $\rho>0$ demands $0<k<\frac{\gamma+2}{3}$. If the cosmic
evolution is amenable to the desirable properly of dimensional
reduction it further
restricts k as $\frac{2\gamma+1}{3}<k<\frac{\gamma+2}{3}$.\\

\textbf{Accelerating universe - I}\\

As discussed earlier in the introduction that our universe is
presently accelerating and at the early phase it was decelerating
. The early deceleration is physically relevant in the sense that
it allows structure formation while the present day acceleration
is in conformity with the current data from the supernovae. So we
now calculate the deceleration parameter for our metric as
\begin{equation}
q = -\frac{\ddot{R}R}{\dot{R^{2}}} =
\frac{\gamma^{2}+(1-3k)\gamma+1}{1-\gamma}
\end{equation}
It is well known that both $\gamma$ and $k$ should be less than
one. But if we consider the condition $0<\gamma<1$, a little
algebra shows that for acceleration $k$ must be greater than 1
which is not desirable. So it is concluded that for power law
expansion  acceleration is not possible for the condition
$0<\gamma<1$ and $\frac{2\gamma+1}{3}<k<\frac{\gamma+2}{3}$.
 Simple calculation shows that acceleration is possible when
$-1<\gamma<-2+\sqrt{3}$ and $-1< k
<\left(\frac{\gamma^{2}+\gamma+1}{3\gamma}\right)$. But in this
case dimensional reduction is not possible. Here three dimensional
scale factor expands indefinitely, matter density is positive and
acceleration is possible. The above features are shown in the
figure 1.
\begin{figure}
  \includegraphics[width=8cm]{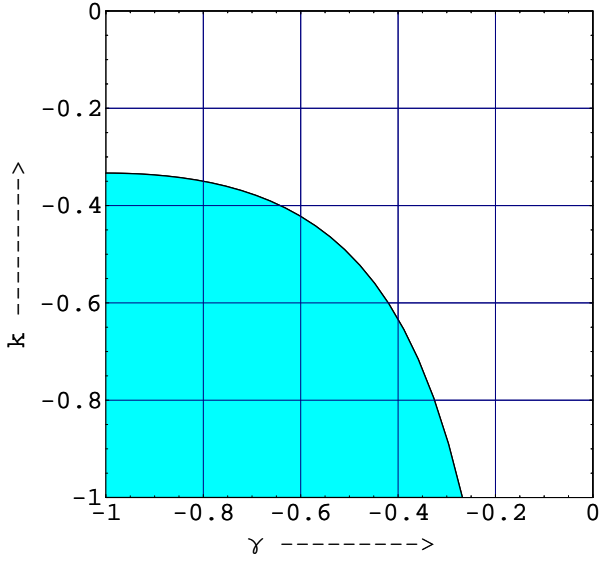}
  \caption{
 \small \emph{The shaded region corresponds to
  the range of $\gamma $ and k where
   universe is accelerating, 3D scale factor is expanding
    and $\rho$ is positive but no dimensional reduction}\label{1}
    }
\end{figure}
If we calculate 4D volume,
\begin{equation}
V = R^{3}A = t^{\frac{-\gamma-3k+4}{\gamma^{2}-3k\gamma+2}}
\end{equation}
it is evident that it expands indefinitely. For a particular value
of $k$ and $\gamma$, all the features are shown in the figure 2.
\begin{figure}
  \includegraphics[width=10 cm]{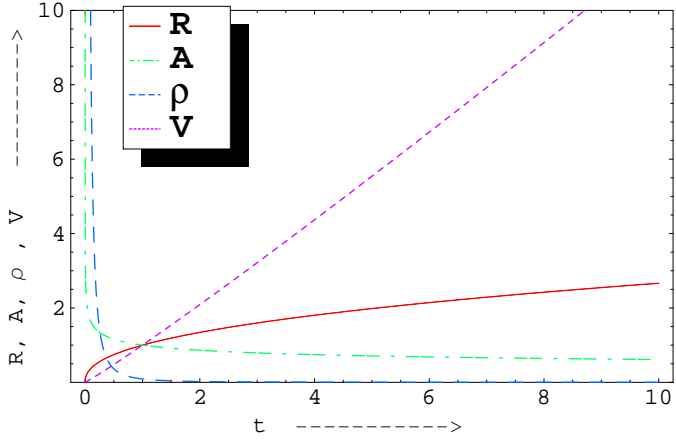}
  \caption{
  \small\emph{The time evolution of R, A, $\rho $ and V are shown in
  this figure taking the positive value of $\gamma $ and k}\label{1}
    }
\end{figure}
At this stage correspondence to some earlier works in this field
may be of interest.\\

1. As discussed earlier when $ m = \frac{1}{2}
$,~~$\rho=p_{5}=p=0$ and we recover the well known solution of
Chodos and Detweiler
\cite{ch} in vacuum when dimensional reduction takes place.\\

2. For $k = \gamma =  0$,  we get the earlier
 solution of Gr\o n \cite{gr} as
\begin{equation}
R = A = t^{\frac{1}{2}}, \\~~ p = p_{5} = 0 ,\\~~ \rho =
\frac{3}{2t^{2}}
\end{equation} \\
where an isotropic expansion in all five dimensions for dust
fluid takes place.\\

3. Next we consider a matter field such that the pressure is
isotropic in all dimensions ( $ k = \gamma $ )
\begin{equation}
R = A = t^{\frac{1}{2(1+k)}}\\,~~ p = p_{5} =
\frac{3k}{2t^{2}(1+k)^{2}}\\,~~
 \rho = \frac{p}{k}
\end{equation}
This is the  homogeneous case of our earlier work \cite{dp}.\\

4. When we consider 3D radiation case, i.e., $ k= \frac{1}{3} $,
we get the following solutions.
\begin{equation}
R = t^{\frac{(1-\gamma)}{(\gamma^{2}-\gamma+2)}} ,~~
 A =
t^{\frac{(2\gamma)}{(\gamma^{2}-\gamma+2)}},~~
 p = \frac{(1-\gamma^{2})}{(\gamma^{2}-\gamma+2)^{2}t^{2}},~~
 p_{5}=
3\gamma p   ,~~
 \rho = 3p
\end{equation}
for $0<\gamma<1$, three dimensional scale factor expands
indefinitely, but dimensional reduction is not possible for extra
space. $\rho$ is also positive for this condition. Again for
$-1<\gamma<0$ dimensional reduction is possible and  $\rho>0$ in
this condition. But extra dimensional pressure will be negative.
No acceleration is possible here, i.e., $q>0$. R expands
indefinitely.\\

5. For $ k = \gamma = -1 $ , we can not get the solutions simply
by putting these values in the expressions (13-17). The field
equations are separately solved for such a case and the solutions
are given by

\begin{equation}
R = A = e^{mt} , ~~\rho = p = p_{5} = 6m^{2}
\end{equation}
The above solutions are same as homogeneous case of our earlier
work \cite{dp}. To sum up,  these solutions resemble some special
 cases of \cite{is}, \cite{ag}\\

 \section*{4. Mohammedi's work }
 In  a recent communication Mohammedi \cite{mh} put forward a new
 proposal regarding the interpretation of matter field as also of the
 higher dimensional equations of motion. Closely following his
 arguments we have in this section presented and discussed a few
 more results. But before presenting our work proper we would like to
 very briefly summarise his main results skipping intermediate mathematical
 details.
 The cardinal point in Mohammedi's work is that if one assumes
\emph{ apriori} a relationship between the 3D scale factor and the
extra scale as
\begin{equation}
\frac{\dot{A}}{A} = -n\frac{\dot{R}}{R}
\end{equation}
( n is an arbitrary constant) the FRW equations of standard 4D
cosmology are obtained precisely through defining a new term what
he calls an\emph{ effective pressure} expressed in terms of the
components of the higher dimensional energy momentum tensor. Thus
one defines the effective pressure in such a way that the higher
dimensional equations of motion yield the usual equations of
motion of ordinary 4D cosmology. The other remaining equation
simply determines, in terms of the radius of our universe, the
pressure along the extra dimensions. It is evident from the the
generalised field equations(5-6) that the expressions between the
curly brackets for $\rho$ and $p$ are similar to the analogous 4D
FRW equations. Therefore the 4D energy density $\rho_{4}$ and the
pressure P may be identified with the quantities $(\rho-
\rho_{0})$ and $(p- p_{0})$ where $\rho_{0}$ and $p_{0}$
respectively denote the terms containing the higher dimensional
coefficients in equations(5-6). This, according to Mohammedi, is
the usual and standard interpretation of higher dimensional
equations of motion. In this interpretation,however, the
$\rho_{4}$ and $P$ contain contributions involving the scale
factors A and R. Then he went on to explore if there exists other
interpretations of the higher dimensional equations of motion and
given the ansatz (24) he claims to find the answer in the
affirmative.\\ To make things more transparent let us compare the
Bianchi identities in the 4D and (4+d)-dim. cases as
\begin{equation}
\frac{d}{dt}\left(R^{3}\rho_{4}\right) +
P\frac{d}{dt}\left(R^{3}\right)=0
\end{equation}
\begin{equation}
\left\{\frac{d}{dt}\left(R^{3}\rho\right) +
p\frac{d}{dt}\left(R^{3}\right)\right\} +
dR^{3}\frac{\dot{A}}{A}\left( \rho + p_{d}\right)=0
\end{equation}
where $\rho_{4}$ and $P$ are the 4D density and pressure
respectively in a 4D FRW space time while $\rho$ and $p$ are the
analogous terms in higher dimensional cosmology. Again the
quantities within the curly brackets in (26) are of the form of 4D
conservation equation (25). Now with the ansatz (24) a little
algebra shows that the last equation can be written as
\begin{equation}
\frac{d}{dt}\left(R^{3}\rho\right) +
\overline{p}\frac{d}{dt}\left(R^{3}\right)=0
\end{equation}
 where $\overline{p}$ is an effective pressure given by
\begin{equation}
\overline{p} = p - \frac{dn}{3}(\rho + p_{d})
\end{equation}
The higher dimensional conservation equation is now exactly of the
same form as that of the $4D$.\\
Now using the ansatz (24) the field equations (4-5) finally reduce
to

\begin{eqnarray}
\frac{\rho}{2\beta} &=&\frac{1}{2}\left[6+ dn
(dn-n-6)\right]\frac{\dot{R^{2}}}{R^{2}}+ \frac{3K}{R^{2}}-\lambda
+\frac{1}{2}\kappa d(d-1)R^{2n}\\
\frac{p}{2\beta} &=& (dn-2)\frac{\ddot{R}}{R}-\frac{1}{2}[2+ dn
(dn+n-2)]\frac{\dot{R^{2}}}{R^{2}}-\frac{K}{R^{2}}+\lambda
-\frac{1}{2}\kappa d(d-1)R^{2n}\\\nonumber
 \frac{p_{d}}{2\beta}
&=& (dn-n-3)\frac{\ddot{R}}{R}-\frac{1}{2}[6+n(d-1)(dn-4)]
\frac{\dot{R^{2}}}{R^{2}}
 -\frac{3K}{R^{2}} +\lambda \\
 &-&\frac{1}{2}\kappa(d-2)(d-1)R^{2n}
\end{eqnarray}

Using the last three equations the effective pressure comes out to
be
\begin{eqnarray}
\frac{\bar{p}}{2\beta} &=&
-\frac{1}{3}[6+dn(dn-n-6)\frac{\ddot{R}}{R}-\frac{1}{6}[6+dn(dn-n-6)]
\frac{\dot{R^{2}}}{R^{2}}-\frac{K}{R^{2}}+\lambda \nonumber\\
&-&\frac{1}{6}\kappa d(2n+32)(d-1)R^{2n}
\end{eqnarray}
Now for realistic cases the terms proportional to $R^{2n}$ have to
absent because matter field can not increase in an expanding
universe forcing us to choose either $d=1$ or a Ricci flat extra
space defined as $\kappa=0$. Let us now therefore choose $\kappa=0$
 ( if d =1, $\kappa =0$ automatically)\\
If one now makes the identifications that
 \begin{eqnarray}
\alpha &=& \frac{\beta}{6}[6+ dn(dn-n-6)]v^{(d)}\\
 k &=& \frac{6K}{[6+dn(dn-n-6)]}\\
\Lambda &=& \frac{6\lambda}{[6+dn(dn-n-6)]}
\end{eqnarray}
where $\alpha=\frac{1}{16\pi G}$ is the 4D gravitational coupling
constant, $\Lambda$ is the 4D cosmological constant and $v^{d}$ is
the finite volume of the extra dim. manifold introduced for
dimensional consideration. Here we asuume that $
[6+dn(dn-n-6)]\neq 0$. But when it is negative the signs of 4D
quantities $k$ and $\Lambda$ are opposite to those of $K$ and
$\lambda$. The 4D quantities $\rho_{4}$ and P are then identified
with
 $\rho_{4}= v^{(d)}\rho$ ,~~~ $P=v^{(d)}\bar{p}$.
Using the above equations we finally get, analogous to the 4D case
the following well known relation ($\Lambda = 0$)
\begin{equation}
6\frac{\ddot{R}}{R} +\frac{1}{2\alpha} ( \rho + 3 \overline{p}
)v^{(d)} = 0
\end{equation}
For accelerating model, $ \overline{p} < - \frac{\rho}{3} $
implying
\begin{equation}
p < ( dn - 1 )\frac{\rho}{3} + \frac{dn}{3} p_{d}
\end{equation}
So the nice thing about the whole analysis is that both $\rho$ and
$p_{d}$may be physically realistic obeying all energy conditions
but only the effective four dimensional pressure, $ \overline{p} $
is negative. In analogy with the curvature quintessence this
ansatz may be termed `dimension driven quintessence'. Assuming  as
before $ p_{5}= \gamma \rho $ the equations (29) and (31) yield
for a 5D (d=1) case
\begin{equation}
\frac{\ddot{R}}{R} + ( \gamma - n \gamma +1)
\frac{\dot{R^{2}}}{R^{2}} + \frac{K(1+\gamma)}{R^{2}} = 0
\end{equation}
The equation (38) simplifies via the transformation
\begin{equation}
\eta = R ^{(\gamma - n\gamma +2)}
\end{equation}
to
\begin{equation}
\ddot{\eta} + K ( \gamma + 1) ( \gamma - n\gamma + 2)
\eta^{\frac{\gamma - n\gamma}{\gamma - n\gamma +2}} = 0
\end{equation}
Multiplying by $\dot{\eta}$ the above gives a first integral as
\begin{equation}
\dot{\eta^{2}} + K (\gamma - n\gamma +2)^{2} \eta^{\frac{2\gamma -
2n\gamma +2}{\gamma - n\gamma +2}} \frac{\gamma +1}{\gamma -
n\gamma +1} = b
\end{equation}
where b is an integration constant. It is not possible to get a
general solution of this equation. However several possibilities
present itself:\\
\textbf{Case I }( b = 0 ) \\
Hence the equation (41) integrates to
\begin{equation}
R = \sqrt{\frac{-K(\gamma + 1)}{\gamma - \gamma n + 1}}~t
\end{equation}
Evidently $K < 0$, pointing to an open 3D space with zero
deceleration parameter. This is Milne's model and has important
astrophysical consequences as discussed by Riess\cite{riess} in
the context of interpreting the findings of high redshift
supernovae for this `coasting universe'. Moreover a little algebra
shows
\begin{equation}
\rho = -\frac{3n}{(1+ \gamma)t^{2}}~~  , ~~   \bar{p} =
\frac{n}{(1+ \gamma) t^{2}}
\end{equation}
with an equation of state  $\bar{p} = - \frac{\rho}{3}$. Moreover
the positivity of energy density implies that n should be
negative. So there will be no dimensional reduction in this case.
This equation of state is only to be expected because
$\ddot{R}= 0$ dictates that $ ( \rho + 3\bar{p}) = 0$.\\
\textbf{Case II} ( $b \neq 0$)\\
To make the equation (41) mathematically tractable let us assume
that the exponent of $\eta$ in this equation is unity i.e.
$\gamma(1- n )= 0$, implying that either $ n=1~~ or~~
\gamma = 0$\\
Taking $n=1$ we get from equation (41)
\begin{equation}
R^{2}= -K(\gamma +1)t^{2}+ lt + m
\end{equation}
where l and m are constants of integration. This equation reduces
to the earlier solution of Mohammedi for the special case of
$\gamma =0$, i.e vanishing 5D pressure. Accelerating model is
possible only if $K < 0$, which however makes the energy density
negative.\\
\textbf{Case III }\\
On the other hand the equation (41) yields a very simple solution
for the special case of $ \gamma=-1$ i.e. $p_{5}= -\rho$. In this
case we get
\begin{equation}
\eta = at
\end{equation}
which , via equation (39) finally gives
\begin{equation}
R = at^{\frac{1}{1+n}}
\end{equation}
With this value of R we finally get
\begin{equation}
\rho = 3\frac{1-n}{(1+n)^{2}}\left(\frac{1}{t^{2}}\right) +
\frac{3K}{R^{2}}
\end{equation}
\begin{equation}
\bar{p}=
\frac{(1-n)(2n-1)}{(1+n)^{2}}\left(\frac{1}{t^{2}}\right)-
\frac{K}{R^{2}}
\end{equation}
Moreover the deceleration parameter takes a very simple form as
$q=n$. So for positive value of $n$ the model is decelerated.
However, for $n<0$ an accelerated expansion results. Evidently for
$n<0$ no dimensional reduction is possible. Incidentally the large
extra dimension\cite{csaki} is not such a bad news these days as
in the past in the context of currently fashionable different
brane inspired models and their quest to resolve the hierarchy
problem in field theory. In fact the prospect of observing these
large extra dimensions by upcoming experiments has of late created
much excitement among
experimentalists\cite{hewett}.\\
It also follows from equations (47-48) that
\begin{equation}
\rho + 3\bar{p} = \frac{6n(1+n)}{(1+n)^{2}}
\end{equation}
such that $n =0$ corresponds to $(\rho + 3\bar{p}) = 0$, which is
simultaneously the condition for $q=0$ as is evident from equation
(43). It may not be  out of place, at this stage, to digress a
little and to refer to a recent and very elegant version of higher
dimensional theory  formulated and developed by Wesson and his
collaborators\cite{wes} according to which in a 5D spacetime when
the metric coefficients depend also on the extra coordinate it is
possible to interpret most properties of matter as a result of 5D
Riemannian geometry. It essentially differs from what Mohammedi
calls the \emph{standard} interpretaion in that here the 5D
spacetime is vacuum such that the 5D Einstein tensor for the
apparent vacuum $ G_{AB}=0$  contain the 4D Einstein's equations $
G_{ij}= T_{ij}$ as a subset with an induced energy momentum tensor
$T_{ij}$ with classical properties of matter. In fact it follows
from the theorem of Campbell that any analytic N-dim. Riemmanian
manifold can be locally embedded in an (N + 1)D Ricci flat
Riemmanian manifold
\cite{psw}\\
Though not exactly similar mention may also be made to an earlier
work of Frolov et al \cite{fr}where a higher dimensional scenario
with some bulk matter content in extra dimensions in a brane
inspired cosmology is discussed and the effective energy tensor
corresponding to what they termed \emph{shadow matter} is
calculated. They went on to show that there exists regions on the
brane where a brane observer notices an apparent violation of
energy conditions (negative pressure and even negative  energy
density). This concept of shadow matter may be of some relevance
for the  effective 4- dimensional equations of state responsible
for the acceleration.
\section*{5. Accelerating Universe - II }
 In the last section, we have shown that in the extra dimensional
 cosmology one can, in principle at least, achieve acceleration with
 regular matter field subject to the fact that the effective pressure
 should be negative. However, the need for structure formation demands
  that there should be an early deceleration followed by a late
 acceleration. We now present a  model which has this combined property.\\
Here we have five unknowns $( A, R, \rho, p, p_{5})$ with three
independent equations. So we assume $p = p_{5}$ such that the
field equations(K=0) give
\begin{equation}
\ddot{A}+ 2\frac{\dot{R}}{R}\dot{A} - \left(\frac{\ddot{R}}{R} + 2
\frac{\dot{R^{2}}}{R^{2}}\right) A = 0
\end{equation}
Incidentally $R=A$ is a particular solution of this equation. To
get a more general solution we substitute $ A=Ru(t)$ to get
\begin{equation}
R\ddot{u}+ 4\dot{R}\dot{u}= 0
\end{equation}
such that
\begin{equation}
\dot{u}= \frac{\beta}{R^{4}}
\end{equation}
where $\beta$ is an arbitrary constant. We make a further ansatz
in the expression of deceleration parameter as
\begin{equation}
  q = \frac{a-R^{m}}{b+R^{m}}
\end{equation}
where a, b and m are arbitrary constants.
Using the usual expression of the deceleration parameter straight
forward integration shows that
\begin{equation}
 R=R_{0}sinh^{n}\omega t
\end{equation}
($n = 2/m $) will be a solution to this equation. Using equation
(18) we get
\begin{equation}
q = \frac{1-n~cosh^{2}\omega t}{n~cosh^{2}\omega t}
\end{equation}
showing that the exponent n determines the evolution of q. While
for $ n > 1$, it is only accelerating but for $ n < 1 $ we are
able to achieve the desirable feature of \emph{flip}, although it
is not obvious from our analysis at what value of redshift this
flip occurs. One can calculate A for different values of n. Taking
$n = 1/4$ we further get
\begin{equation}
    A = sinh^{\frac{1}{4}}\omega t \left( \beta \ln \tanh \frac{\omega
    t}{2}+\gamma \right)
\end{equation}
where $\alpha $, $\beta$, $\gamma$ and $\omega $ are constants.
From equation (26) and (27) we obtain,
\begin{eqnarray}
  p = p_{5} = - \frac{3\omega^{2}}{8 \sinh^{2}\omega t}( \sinh^{2}\omega t - 1) \\
   \rho = \frac{3 \omega^{2}}{8}\frac{\cosh \omega t}{\sinh^{2}\omega
   t}\left(\cosh \omega t + \frac{2 \beta}{\beta \ln \tanh \omega t   +
   \gamma}\right)
\end{eqnarray}
Depending on the signature and relative magnitudes of the
arbitrary constants the fifth dimension either expands
indefinitely or collapses in a finite time.\\
Interestingly the deceleration parameter is found to be
\begin{equation}
    q = \frac{3 - sinh^{2}\omega t}{1 + sinh^{2} \omega t}
\end{equation}
such that an initially decelerating model starts accelerating
after the critical time $t = t_{c}$ given by  $ sinh \omega t_{c}
> \sqrt{3} $ ( see figure 3 )
\begin{figure}
  \includegraphics[width=10cm]{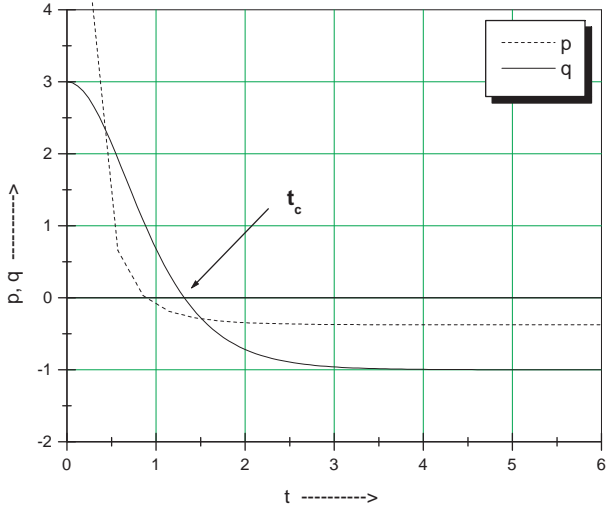}\\
  \caption{
  \small\emph{The time evolution of p and q is shown in this figure.
  When $ t > t_{c}$ q becomes negative. }\label{1}
    }
\end{figure}
 but pressure starts becoming negative earlier at
$sinh \omega t > 1$. This is not surprising. There is plenty of
observational evidence for a decelerating universe in the recent
past\cite{agr}, \cite{turner}. But the dominance of negative
pressure does not guarantee the present acceleration of the
universe. For the universe to accelerate the the negative pressure
has to dominate long enough as to overcome the gravitational
attraction produced
by ordinary matter\cite{ponce}.\\
\textbf{Special case} ( A= R)\\
We have mentioned earlier that $A= R$ is a particular solution of
the equation (50). Though simple, in what follows, we shall
presently see that this choice is rich with various possibilities
in interpreting our matter field as also in comparing the evoluon
of
 the universe with a Chaplygin type of fluid.\\
With $ A = R = sinh^{n}\omega t$ we get
\begin{equation}
\rho = 6n^{2}\omega^{2}+ \frac{6n^{2}\omega^{2}}{sinh^{2}\omega~t}
= \Lambda + \frac{B}{R^\frac{2}{n}}
\end{equation}
\begin{equation}
p = - \Lambda - \frac{2n - 1}{2n} \frac{B}{R^\frac{2}{n}}
\end{equation}
where $\Lambda = 6n^{2}\omega^{2}$ and $B = \Lambda
R_{0}^{\frac{2}{n}}$. One might recall that from equation (4) with
$ R=A$ and $\gamma = k$ it follows that for, $p = k \rho$ we get
\begin{equation}
\rho = R^\frac{-1}{4(1+k)}
\end{equation}
such that, $n=\frac{1}{4}$ corresponds to a stiff fluid ( k=1)
with $\rho \sim \frac{1}{R^{8}}$ and $n=\frac{2}{5}$ to a
radiation dominated phase $( k=\frac{1}{4}) $ with $\rho \sim
\frac{1}{R^{5}}$ and lastly $n=\frac{1}{2}$ to a matter dominated
model(k=0) with $\rho \sim \frac{1}{R^{4}}$. Thus, interesting to
point out that the exponent $n$ in equation (64 ) characterises
the nature of the fluid we are dealing with. We can make the
identification clearer if we write a sort of equation of state
using equations (60- 61)as
\begin{equation}
p = \frac{1 - 2n}{2n}\rho - \frac{\Lambda}{2n}
\end{equation}
 such that $ k = \frac{1 - 2n}{2n}$. One can now identify the two
 arbitrary constants in (54) as $ \omega = \sqrt{\frac{2
 \Lambda}{3}}( k + 1)$ and $ n = \frac{1}{2(k+1)}$ such that we
 finally get
\begin{equation}
R = R_{0}[\sinh\sqrt{\frac{2\Lambda}{3}}( k + 1) t
]^{\frac{1}{2(k+1)}}
\end{equation}
So we get the following cases of matter field \\
a. $(n = \frac{1}{4})$ ( stiff fluid)
\begin{equation}
\rho = \Lambda + \frac{B}{R^{8}} ~~and~~ p = -\Lambda +
\frac{B}{R^{8}}
\end{equation}\\

b. $(n=\frac{2}{5})$ (radiation)
\begin{equation}
\rho = \Lambda + \frac{B}{R^{5}} ~~and~~ p = -\Lambda +
\frac{B}{4~R^{5}}
\end{equation}\\

c. $(n=\frac{1}{2})$ ( dust)
\begin{equation}
\rho = \Lambda + \frac{B}{R^{4}} ~~and~~ p = -\Lambda
\end{equation}
We see that for small R the equation in \emph{case a} is
approximated by $\rho = \frac{\Lambda}{R^{8}}$, which corresponds
to a universe dominated by a stiff fluid in 5D spacetime.
Similarly the \emph{case b} and\emph{ case c} refer to radiation
dominated and dust dominated universe respectively. On the other
hand for a large value of the cosmological radius we see that the
above equations suggest that $\rho = \Lambda ~~ and~ p =-
\Lambda$~ which, in turn, corresponds to an empty universe with a
cosmological constant $\Lambda $ ( i.e., a de Sitter
universe)\\
Thus equations (63-65) describe the mixture of a cosmological
constant with a type of fluid obeying some equation of state. The
last case known as 'stiff fluid` characterised by the equation of
state, $ p = \rho$ is particularly interesting. Note that a
massless scalar field is a particular instance of stiff matter.
Therefore, in a generic situation, our cosmology may be looked
upon as interpolating between different phases of the universe
from a stiff fluid, radiation or dust dominated universe to a de
Sitter one passing through an intermediate phase which is a
mixture just mentioned above. The interesting point, however, is
that such an evolution may be accounted for by using one fluid
only as opposed to the earlier works  \cite{gorini},\cite{barrow}
representing simple two fluid model. Correspondence to models
driven by a generalised Chaplygin type of fluid \cite{anjan}
described by an equation of state
\begin{equation}
\rho = \left(\Lambda +
\frac{B}{R^{3(1+\alpha)}}\right)^{\frac{1}{1+\alpha}}
\end{equation}
 is only too  apparent although here, as mentioned before we do not need to
hypothesise the existence of a mysterious type of fluid to explain
the observations. Here $\alpha $ is an additional free parameter
to play with to fit the observational data. In the light of above
discussions the behaviour of the deceleration parameter in our
model is as expected (fig.3). Initial dust dominance provides the
gravitational pull for the expansion to decelerate but once the
cosmological term starts dominating acceleration occurs with, $q =
-1$. Although evolution of this kind has been exhaustively
discussed in the literature in 4D space time but we are not aware
of models of similar kind in higher dimensional spacetime.
Moreover we know that for a sheer-free evolution, if the temporal
dependence of the scale factor is given, one can construct a
potential for a minimally coupled scalar field which would
simulate the evolution as with a perfect fluid. Let us illustrate
the situation in our model. For the Lagrangian
\begin{equation}
L(\phi) = \frac{1}{2}\dot{\phi^{2}} - V(\phi)
\end{equation}
we get the analogous energy density as
\begin{equation}
\rho_{\phi} =\frac{1}{2}\dot{\phi^{2}} + V(\phi) = \Lambda +
\frac{B}{R^{\frac{2}{n}}}
\end{equation}
and the corresponding 'pressure` as
\begin{equation}
p_{\phi} =\frac{1}{2}\dot{\phi^{2}} - V(\phi) =- \Lambda -
\frac{B}{R^{\frac{2}{n}}} + \frac{B}{2n~R^{\frac{2}{n}}}
\end{equation}
such that
\begin{equation}
\dot{\phi^{2}} =\frac{B}{2n~R^{\frac{2}{n}}}
\end{equation}
which, in turn, gives via equation (5) for flat 4D space
\begin{equation}
\phi' = \sqrt{\frac{3B}{n}}\frac{1}{R \sqrt{\Lambda
R^{\frac{2}{n}}+B}}
\end{equation}
where $\phi'$ denotes differentiation w. r. t. the scale factor R.
Integrating we get,
\begin{equation}
\phi = \sqrt{\frac{3n}{4}}\ln \frac{\sqrt{B} - \sqrt{\Lambda
R^{\frac{2}{n}}+B}}{\sqrt{B} + \sqrt{\Lambda R^{\frac{2}{n}}+B}}
\end{equation}
Using equation (54) we finally get
\begin{equation}
\phi = \sqrt{3 n}\ln \tanh\frac{\omega t}{2}
\end{equation}
On the other hand simple algebra shows that
\begin{equation}
V ( \phi ) = \Lambda \left( 1 + \frac{4 n -1 }{4 n}
\sinh^{2}\frac{\phi}{\sqrt{3n}}\right)
\end{equation}
For the dust case ( $n = \frac{1}{2}$ )
\begin{equation}
V ( \phi ) = \Lambda \left( 1 + \frac{1}{2 \sinh^{2} \omega t}
\right)
\end{equation}
while for the analogous stiff fluid case ( $n = \frac{1}{4}$ )
yields a constant potential $ V(\phi ) = \Lambda = V_{0} $ \\
It may not be out of place to call attention to a quintessential
model driven by a tachyonic scalar field \cite{gorini} with a
potential in 4D space time
\begin{equation}
V ( T ) = \frac{\Lambda}{\sin ^{2} \left( \frac{ 3 \sqrt{\Lambda (
1+k )}}{2}\right) T}\sqrt{1 - ( 1+ k) \cos^{2}\left(\frac{ 3
\sqrt{\Lambda ( 1+k )}}{2}\right) T}
\end{equation}
(T is a tachyonic scalar field) giving the cosmological evolution
as
\begin{equation}
R (t) = R_{0} \left( \sinh \frac{3 \sqrt{\Lambda} ( 1+ k )
t}{2}\right)^{\frac{2}{3 ( 1+ k)}}
\end{equation}
It behaves like a two fluid model where one of the fluids is a
cosmological constant while the other obeys a state equation $ p =
k \rho $, $(-1 < k < 1 )$.  Similarity of this evolution with our
model is more than apparent except for some numerical factors
coming out because we are  here dealing with a higher dimensional
spacetime. But the main result may be re-emphasized that we get
this evolution without forcing ourselves to invoke any extraneous
tachyonic type of scalar field. To end the section a final remark
may be in order. From the equation (63) it follows such that the
sound speed is given by
\begin{equation}
C_{s}^{2} = \frac{\delta p}{\delta \rho}= \frac{1-2n}{2n}
\end{equation}
which implies that to avoid imaginary value of the speed of sound
$n< \frac{1}{2}$. Evidently in the dust model $(n=\frac{1}{2})$
$C_{s}$ vanishes as expected. This along with the requirement that
$C_{s}$ should never exceed the speed of light further restricts
the range of $n$ as $ \frac{1}{4}< n < \frac{1}{2}$. \\

Before concluding the section we call attention to a serious
defect of the present analysis. Here we have postulated a 5D
matter field. But what is relevant is the effective 4D physical
quantities. In line with our discussions in section 4 we can
calculate the 4D quantities as $(\rho-\rho_{0})$ and $(p - p_{0})$
in this case also. From equation (5) it follows that with $R=A$
and $d=1$, $\rho_{0}= 3\frac{\dot{R^{2}}}{R^{2}}$ such that the
expression $(p - p_{0})$ turns out to be qualitatively of the same
form as in equation (60). Only the numerical factor differs. So
most of our findings remains essentially unaltered, which is
hardly surprising because with $R=A$ the radial and the extra
fifth coordinate are exactly equivalent. So we are brief on this
point.
\section*{6. Discussion }
In this work we have discussed a 5D homogeneous model with
maximally symmetric 3D space. As the field equations are under
determined we are forced to  assume two equations of state
connecting pressure and density. But it should also be emphasised
that,for the sake of mathematical simplicity, we have chosen to
sacrify some generality and to assume a number of relations in
sections 4 and 5 to make the field equations integrable.
Nevertheless our solutions are quite general in nature in the
sense that many well known results in this field are recovered as
special cases. Fixing the magnitude of the arbitrary constants we
have ensured the positivity of the matter field, good energy
conditions as well as dimensional reduction. We have taken only
one extra spatial dimension but we believe most of the findings
may be extended if we take a larger number of extra dimensions.
The most important finding in this work, in our opinion, may be
summarised as : we do not have to hypothesise the existence of an
extraneous scalar field with mysterious properties of matter to
achieve an accelerating universe. However one should admit that
here we have to postulate the existence of an extraneous 5D matter
field instead as also some other assumptions to achieve the
acceleration. Relevant to point out that in an interesting work Li
Quiang etal \cite{li} also recently showed that the Brans Dicke
theory generalised to five dimensions is reduced to a 4D theory
where the 4-metric is coupled to two scalar fields, which may
account naturally for the present accelerated expansion of the
universe. The extra matter field in our model is of geometrical
origin which is, however, not very uncommon in the literature.
Correspondence to curvature quintessence, Wesson's induced matter
theory as also the shadow matter concept of Frolov etal in the
context of brane cosmology may be of some relevance here. To end
the section we like to point out some serious shortcomings of our
model which need considerable refinements in future exercise. In
section 5 the model is based on assumption of a specific form of
the deceleration parameter, which definitely suffers from the
disqualification of a sort of ad-hocism. Moreover the proper
interpretation of matter field in higher dimensional models
continues to plague the workers in this field. In the cosmological
context one starts with a 5D matter field and looks for various
types of dynamical compactification of the extra dimensions.It is
conjectured that some stabilising mechanism ( quantum gravity may
be a potential candidate) should finally halt the continual
shrinkage such that it stabilises at a planckian length so as to
be unobservable with the low energy physics available today. So
with this phase transition the extra metric coefficients lose
their dynamical character and the field equations along with the
matter field are effectively four dimensional and it enters
exactly the 4D FRW phase. So in this scenario there is no
effective 4D properties of matter. While some of the solutions in
section 3 are amenable to the desirable feature of dimensional
reduction in section 5 we have taken the fifth dimension in the
same footing with the rest as $R=A$. So dimensional reduction is
clearly absent. Apart from this undesirable feature a serious
defect of our work is the absence of any stabilising mechanism
itself which should finally halt the continual shrinkage of the
extra space. In this regard comparison with analogous curvature
driven quintessence models are striking where most of the
solutions are unstable against perturbations. In that context
Guendelman and Kaganovich \cite{gk} showed earlier that
Wheeler-deWitt equation in ADS space time does provide a quantum
repulsive effect to stabilise the extra spatial volume. It is also
shown that if one works with more than one extra dimension it may
create a repulsive potential to avoid the singularity of zero
extra spatial volume \cite{zhuk}
,\cite{zh}.\\
To conclude a final remark may be in order. In section 5 the
assumtion $R=A$ generates a matter field which may be
interpretated as a mixture of perfect fluid obeying an equation of
state as well as a cosmological constant with either term
dominating at different phases of evolution allowing a smooth
transition from a decelerating to an accelerating model. With no
dynamical reduction this form of matter field is open to serious
criticism as we no longer recover the 4D cosmology nor the
effective 4D properties of matter. But both in Wesson's STM theory
or its equivalent brane models \cite{leon} it is the effective 4D
physics that matters.In fact the acceleration is here made
possible because we have introduced a 5D matter at the expense of
an extraneous scalar field with peculiar properties. Further when
one imposes the cylindricity condition of Kaluza-Klein the induced
matter in the STM theory is either radiation-like or empty
\cite{poncedeleon}, which certainly can not be the source of
acceleration.

As a future exercise one should envisage an additional scenario
with other inputs such that the currently observed acceleration is
followed by a decelerating phase, which finally hits a big brake
singularity.

\textbf{Acknowledgment : } S.C. wishes to thank TWAS, Trieste for
travel support and ITP (Beijing) for local hospitality where the
work was initiated while DP acknowledges financial support of UGC,
New Delhi. We also thank the anonymous referee for comments which
led to a significant improvement of the earlier version.


\begin{thebibliography}{22}
\bibitem{pd} T. Padmanabhan  - `Understanding our
Universe : Current status and open issues'  and references
therein, gr- qc / 0503107
\end{thebibliography}
\end{document}